\documentclass[a4paper]{revtex4}
\begin{document}
\title{Identical particles are indistinguishable but...}
\author{A. C. de la Torre}\email{delatorre@mdp.edu.ar}
 \author{H. O. M{\'a}rtin}
 \affiliation{ IFIMAR - CONICET \\ Departamento de F{\'\i}sica -
 Facultad de Ciencias Exactas y Naturales -
 Universidad Nacional de Mar del Plata\\
 Funes 3350, 7600 Mar del Plata, Argentina.}
\begin{abstract}
It is shown that quantum systems of identical particles can be
treated as if they were different when they are in well
differentiated states. This simplifying assumption allows the
consideration of quantum systems isolated from the rest of the
universe and justifies many intuitive statements about identical
systems. However, it is shown that this simplification may lead
to wrong results in the calculation of the entropy.
\end{abstract}
\maketitle
\section{INTRODUCTION}
At the site \emph{http://www.sandia.gov/surface\_science/stm/}
one can see moving pictures, taken with tunnel effect microscope,
of \emph{Si} dimers (two \emph{Si} atoms stick together) moving
in a substrate. The dimers diffuse on the surface in one
preferred direction and sometimes get trapped between two fixed
chains of \emph{Si} atoms and move back and forth. Every common
sense physicist would agree with the sentence: \emph{``the dimer
moves back and forth within the trap''}. However, a more careful
analysis reveals that the sentence is in contradiction with the
principles of quantum mechanics because when we say ``the dimer''
we are assigning an identity to an indistinguishable particle.
There are many cases where we use \emph{different states} in
order to distinguish or identify particles that are
indistinguishable. This is clearly an error violating a well
established principle of quantum mechanics but we may ask how
serious this error is ``for all practical purposes''
(FAPP)\cite{bell}. After all, when we consider an isolated
system, like a hydrogen atom, we are identifying it among myriads
of other identical systems and we never question this violation
of quantum mechanics. When I think of an individual electron
hitting the screen of my PC and producing light as a particle
different from another electron hitting the referee's computer
screen, I am in a contradiction with quantum mechanics. The same
error I make when I consider one \emph{Si} atom in a chip of my
computer as different from another \emph{Si} atom in a chip of
his computer. How serious is this? Strictly speaking, we are
almost always dealing with physical systems build with identical
particles and in principle we should treat them as dictated by
quantum mechanics, that is, with states symmetric or
antisymmetric under the permutation of particles. As far as we
know today, the only different particles are the three fermion
families of quarks and leptons and the gauge bosons of the
interactions but in the future we may discover that they are also
just different states of some unique particle or system (string,
brane, or whatever).

The universe is built of a large number of one or a few identical
particles and quantum  mechanics teach us how to treat them
accordingly. Of course it would be a big nuisance to consider
every physical system as an indistinguishable part of the whole
universe and therefore it is important to show that, FAPP, the
treatment of identical systems, when they are in well
differentiated states, can be done as if they were different
particles. It has been rigorously proved\cite{herbut} that the
quantum system of two identical particles in different states is
equivalent to the corresponding system of two different particles
and we are therefore allowed to treat identical particles as
different when they are in well differentiated states, for
instance, far away or with clearly different properties. The two
particle system contains the main ingredients of the problem and
the results obtained with it can be generalized to $N$ particles
system. The proof mentioned above is rigourous but its
``readability'' can be substantially improved. We will therefore
see here a simpler presentation of these results in a way
accessible to students but without loss of rigour and generality.
Complementing the mathematical proofs, we also include many
discussions and comments related with the general problem of
identical particles in quantum mechanics and in classical physics
with more emphasis in the physics of the problems rather than
with its formal aspects. One important point emphasized in this
work is that the simplifying assumption that sometimes identical
particles can be treated as different,  may lead to wrong results
when calculating the entropy of a system. For the didactic
application of this work we suggest several simple exercises.
\section{TWO PARTICLE FORMALISM}
In order to clarify the notation, we present the Hilbert space
formalism for the quantum mechanical treatment of a two particle
system, identical or different. Let $\mathcal{H}$ be the Hilbert
space of states for the description of a one particle system.
Then, as is well known, the state of a two particle system is an
element of $\mathcal{H}\otimes\mathcal{H}$ (in general, both
spaces need not be of the same dimension but for the cases of
interest in this work the dimension is the same and we don't need
to take different Hilbert spaces in the tensor product). In some
cases, a redundant notation is used
$\mathcal{H}_{1}\otimes\mathcal{H}_{2}$ in order to associate the
first particle to $\mathcal{H}_{1}$ and the second to
$\mathcal{H}_{2}$; we will not do this because the right-left
order of the factors is sufficient to denote the association.

The permutation of the two particles, or equivalent, the
permutation of the states associated to each particle, is an
operation of fundamental importance for the definition of the
concept of identical particles. We define then the
\emph{Permutator} (a unitary transformation) as a linear operator
$\Pi: \mathcal{H}\otimes\mathcal{H}\rightarrow
\mathcal{H}\otimes\mathcal{H} $ by
\begin{equation}\label{permut}
    \Pi\ (\Psi\otimes\Phi) =
\Phi\otimes\Psi\ .
\end{equation}
From this definition, it follows as a simple exercise that
\begin{eqnarray}
  \Pi^{2} &=& \textbf{1}\ , \\
  \Pi\chi &=& \lambda\chi \hbox{ with }\lambda=\pm 1\ , \\
 \Pi^{\dag}&=& \Pi\ , \\
  \Pi^{\dag}( A\otimes B)\Pi &=& B\otimes A \ ,
\end{eqnarray}
where $\chi$ is an eigenvector of $\Pi$ and $A$ and $B$ are
arbitrary operators (the last relation follows easily from the
spectral decomposition of the operators). Now we can state the
fundamental quantum mechanical postulate for identical particles.
\emph{The state of  a system of two identical particles is an
eigenvector of the permutator $\Pi$ with eigenvalue $\lambda=1$
for identical bosons and $\lambda=-1$ for identical fermions.}
This postulate has an intuitive interpretation: imagine that we
try to describe a two identical particle system with a state
$\Psi\otimes\Phi$. Clearly this is inappropriate because we are
identifying the first particle with $\Psi$ and the second with
$\Phi$ in contradiction with the assumption that they are
identical. We must therefore correct this error by adding (or
better, superposing) the opposite association. That is,
$\Psi\otimes\Phi\pm\Phi\otimes\Psi$ which is an eigenvector of
$\Pi$. This is the strongest form of the identical particle
postulate from which other weaker versions can be
deduced\cite{ball,mess}. From this postulate it follows (simple
exercise) that every observable of a two identical particle
system, represented by an operator $O:
\mathcal{H}\otimes\mathcal{H}\rightarrow
\mathcal{H}\otimes\mathcal{H}$, must be invariant under the
permutator. That is, $\Pi^{\dag}  O \Pi=O$ or $[O,\Pi ]=0$. The
two identical particle observables are therefore of the type
$A\otimes B+B\otimes A$. The fact that we deal with identical
particles does not forbid us to consider \emph{one particle}
observables that must take the form $A\otimes
\textbf{1}+\textbf{1}\otimes A$ and therefore we can not use the
measurement of this one particle observables in order to identify
a particle.

The form of the state of two identical particles,
$\Psi\otimes\Phi\pm\Phi\otimes\Psi$, reminds us the
\emph{entangled states} where two subsystems share two properties
but  both properties are shared by both subsystems in a holistic
way without individual assignment. For instance, in the singlet
state of two spin 1/2 particles (identical or not)
$1/\sqrt{2}(\varphi_{+}\otimes\varphi_{-}-\varphi_{-}\otimes\varphi_{+})$
-a paradigmatic example for entangled states- we can not tell
which particle has spin up and which one has spin down. Both
particles share spin up and down simultaneously. A similar
situation appears for the state of two identical particles that
can be thought of as a state of \emph{identity entanglement}.

With the permutator operator $\Pi$ we can define a
\emph{Symmetrizer} $\mathcal{S}$ and an \emph{Antisymmetrizer}
operator $\mathcal{A}$:
\begin{eqnarray}
  \mathcal{S} &=& \frac{1}{2}(\textbf{1}+\Pi) \\
  \mathcal{A} &=& \frac{1}{2}(\textbf{1}-\Pi)\ .
\end{eqnarray}
These two operators are projectors that project in the subspaces
$\mathcal{H}_{S}$ and $\mathcal{H}_{A}$ that are orthogonal
$\mathcal{H}_{S}\perp\mathcal{H}_{A}$ and complete. That is, the
Hilbert space is decomposed as an orthogonal sum
$\mathcal{H}\otimes\mathcal{H}=\mathcal{H}_{S}\oplus\mathcal{H}_{A}$.
(Simple exercise: prove all this). The state of two identical
fermions is an element of $\mathcal{H}_{A}$, the state of two
identical bosons belongs to $\mathcal{H}_{S}$ and the state for
two different particles is in $\mathcal{H}\otimes\mathcal{H}$ and
can be decomposed in a symmetrical part plus an antisymmetrical
part. For this work, where we are comparing identical particles
with different particles, it is irrelevant whether the particles
are bosons or fermions an it is therefore convenient to unify
both projectors above in an \emph{Identical Particle Projector}:
\begin{equation}\label{proj}
  \mathcal{I} = \frac{1}{2}(\textbf{1}+\lambda\Pi)\hbox{ with }
   \lambda=\pm
  1\ .
\end{equation}

\section{DIFFERENTIATING IDENTICAL PARTICLES}
In this section we will show that when two identical particles
are in clearly different states they can be treated as different
particles. Of course, this is not true for all states and
therefore we must clearly state what it means that two states are
differentiating.  Given two possible one particle states $\Psi$
and $\Phi$, we say that they are \emph{differentiating} if
$\langle\Psi,\Phi\rangle=0$. This rigourous criterium can be
weakened to \emph{FAPP differentiating} when
$\langle\Psi,\Phi\rangle\approx 0$, that is, the scalar product
is so small that it will have no measurable consequence. Two
eigenstates corresponding to different eigenvalues of an
observables are differentiating. Two gaussian states such that
their widths are much smaller than their separation are FAPP
differentiating (this can be a model for a \emph{Si} atom in my
computer and a \emph{Si} atom in another computer far away).
Sometimes two states can be differentiating due to some different
internal property as, for instance, for two electron states with
arbitrary location but one with spin ``up'' and the other with
spin ``down'', we have $\langle\Psi(x)\otimes\varphi_{+},
\Phi(x)\otimes\varphi_{-}\rangle= \langle\Psi(x),\Phi(x)\rangle\
\langle\varphi_{+}, \varphi_{-}\rangle=0$. Two photon states
corresponding to orthogonal polarization are differentiating. As
with the \emph{Si} atoms above, all localized states in the
classical limit are FAPP differentiating.

However, differentiating states are not sufficient  in order to
differentiate identical particles. We need more. Besides
differentiating states we need to define also
\emph{differentiating observables}, that is,  observables that
are sensitive to differentiating states. In order to understand
this need, consider a two identical particle observable $A\otimes
B+B\otimes A$. The expectation value of this observable in an
identical particle state, \emph{even if it is a differentiating
state}, will involve both observables $A$ and $B$, both states
$\Psi$ and $\Phi$ but also both particles. All possible
combinations appear. What we need, is to associate one
observable, say $A$, with one state, say $\Psi$, and the other
observable $B$ with $\Phi$. For instance, I may be interested in
the energy observable of a $Si$ atom in \emph{my} computer (and
not in his) and therefore we want to relate the energy observable
with the state $\Psi$ denoting the localization in my computer.
We obtain these state sensitive observables $A\rightarrow
A_{\Psi}\ ,B\rightarrow B_{\Phi}$ using the projectors in the
corresponding states,
\begin{eqnarray}
  A_{\Psi} &=& P^{\dag}_{\Psi}A P_{\Psi}\ , \\
  B_{\Phi} &=& P^{\dag}_{\Phi}BP_{\Phi} \ ,
\end{eqnarray}
where the projector (hermitian and idempotent) in the Hilbert
space formalism is given by $P_{\Psi}= \Psi \langle\Psi ,
\cdot\rangle$ or, for those addict to the Dirac's notation,
$P_{\Psi}= |\Psi\rangle \langle\Psi |$. Physically, the operator
$A_{\Psi}$ corresponds to the simultaneous measurement of the
observable $A$ and all properties characteristic of the state
$\Psi$ and therefore we must have $[A,P_{\Psi}]=0$ and also
$[B,P_{\Phi}]=0$. With this, we have $ A_{\Psi}=A P_{\Psi}$ and $
B_{\Phi}=B P_{\Phi}$. These state sensitive observables are the
quantum mechanical counterpart of the \emph{classical}
observables: the value of an observable in classical physics is
always given by the state of the system, that is, they are
functions of the coordinates and their associated canonical
momenta that determine the state of the system. Notice that the
association of observables with states is perfectly ``legal'' and
it is not a violation of the identity principle: state sensitive
observables for two identical particles are built according to
the rules of quantum mechanics as
\begin{equation}\label{diffobserv}
  A_{\Psi}\otimes B_{\Phi}+B_{\Phi}\otimes A_{\Psi}\ .
\end{equation}

We can now prove that the treatment of two identical particles in
two differentiating states is equivalent to the treatment of two
different particles in the corresponding states in the sense that
the expectation values of every differentiating observable is the
same in both cases.  This is the main result of
reference\cite{herbut}. Let us assume that we have two identical
particles sharing two differentiating states $\Psi$ and $\Phi$;
therefore the state of the system is given by
\begin{equation}\label{idenstat}
   \Xi_{ID} =  \sqrt{2}\ \mathcal{I}(\Psi\otimes\Phi) =
   \frac{1}{\sqrt{2}}(\Psi\otimes\Phi+\lambda\Phi\otimes\Psi) \ .
\end{equation}
On the other side if the particles are different the state is
\begin{equation}\label{difstat}
   \Xi_{DIF} =  \Psi\otimes\Phi \ ,
\end{equation}
(or the opposite association $\Phi\otimes\Psi$). Notice that
there is a one to one relation between $\Xi_{ID}$ and
$\Xi_{DIF}$; in fact, applying $\sqrt{2}\mathcal{I}$ to
$\Xi_{DIF}$ we obtain $\Xi_{ID}$ and the inverse map is
$\sqrt{2}P_{\Psi}\otimes P_{\Phi}$. In spite of this isomorphism,
both states denote clearly differen physical situations: The
state $\Xi_{ID}$ describes a system of two identical particles
where one particle has the properties associated to $\Psi$ and
one particle has the properties of $\Phi$ but there is no
possibility to decide which one is which; whereas the  state
$\Xi_{DIF}$ we have two different particles, the first one
uniquely identified with the properties of $\Psi$ and the second
with the properties of $\Phi$. We can now present a mathematical
equation that relates these two physically different situations:
by direct calculation (exercise) we obtain that for any arbitrary
observables $A$ and $B$
\begin{equation}\label{id.dif}
   \left\langle\Xi_{ID},\ (A_{\Psi}\otimes B_{\Phi}+
   B_{\Phi}\otimes A_{\Psi})\Xi_{ID}\right\rangle =
   \left\langle\Xi_{DIF},\ (A\otimes B)\Xi_{DIF}\right\rangle\ .
\end{equation}
(In FAPP differentiating states we neglect terms of order
$|\langle\Psi,\Phi\rangle|^{2}$ and get an approximation instead
of an equality). Notice that the fact that this equation involves
expectation values is not a restriction because every
experimentally accessible quantity can be given as the
expectation value of a appropriately chosen operator. Therefore,
an identical particle system in any differentiating state can be
treated as a different particle system with regards to every
differentiating observable. With this result, the desired
simplification in all physical situations mentioned in the
introduction are justified.
\section{SHORT VISIT TO ELEMENTARY PARTICLE PHYSICS}
Notice that the equivalence of identical particles in
differentiating states with different particles can be applied in
both directions. We can treat two identical particles in a
simpler way as different particles but also the other way, we can
treat two different particles as identical but in differentiating
states. We would first think that nobody would be interested in
introducing this complication, however, this possibility has
resulted in important discoveries in particle physics. The first
application of this idea is due to Heisenberg\cite{hei} when he
recognized that the two different particles -proton and neutron-
could be treated as the same identical particle, the nucleon, but
in two different states corresponding to another observable that
Wigner\cite{wig} called \emph{Isospin}. In the same way that we
don't consider two electrons with different spin as different
particles (although they behave differently, for instance, in an
inhomogeneous magnetic field) the proton and neutron are the same
identical particle with isospin ``up'' and ``down''. The
discovery of isospin had important application in the study of
the forces that bind the nucleus: these forces are invariant
under rotations in isospin, a fact known as ``charge
independence'' of the nuclear forces. Spin and isospin 1/2 are
the smallest representations of a symmetry group denoted by
$SU(2)$. Particles with tree charge values, for instance
$\pi_{+},\pi_{0},\pi_{-}$ correspond to identical particles with
isospin one, and in the same way, all hadrons could be assigned
an isospin value. The classification of all known hadrons
suggested the introduction of the larger group $SU(3)$ that
contains isospin and another property (strangeness, initially
called ``hypercharge''). In this scheme, for instance, the eight
particles
$p,n,\Sigma_{+}\Sigma_{0}\Sigma_{-},\Lambda,\Xi_{0},\Xi_{-}$
could be considered as one particle in eight different states of
isospin and strangeness. All the then known hadrons could be
assigned to a $SU(3)$ multiplet. The search of the smallest
representation of the group $SU(3)$ led to the discovery of the
tree quarks $u,d,s$. The experimental discovery of new hadrons
that did not fit in the scheme, forced the introduction of larger
groups and led to the discovery of other quarks. We will not give
more details about particle physics in this work. We just wanted
to point out that the idea of treating different particles as
identical ones in different states, had far reaching consequences
in the reductionist study of nature. As mentioned in the
introduction, this reduction to more and more fundamental
particles is perhaps not finished.

\section{IDENTICAL PARTICLES IN CLASSICAL PHYSICS}
The arguments presented in previous sections can be used in order
to understand the classical limit of statistical mechanics. As a
matter of principles, we must start with the idea that all atoms
or particles are identical and therefore, in all rigour, should
be treated as quantum mechanics dictates, with a state obtained
as a superposition of all possible permutations of one particle
states. In the same way, all observables should be invariant
under the transformation implied by each permutation of particles
(as we did for two particles where the state and observables had
the form $\Psi\otimes\Phi\pm\Phi\otimes\Psi$ and $A\otimes
B+B\otimes A$). Now, from this identical particle quantum
statistics we can make a transition to different particles
statistic when all $N$ particles in the system can be assigned to
$N$ mutually FAPP differentiating states $\Psi_{k}\ ,k=1\cdots
N$. For instance, let each $\Psi_{k}$ be gaussian states with
widths much smaller than the average separation (small width,
$\Delta^{2}_{x}\rightarrow 0$, implies large momentum spread,
$\Delta^{2}_{p}\rightarrow \infty$, that is, large $\langle
P^{2}\rangle$, or large mean kinetic energy, that is, high
temperature). As before, we can now associate observables with
states by the definition of the state sensitive observables, for
instance, for the one particle hamiltonian, $H\rightarrow
H_{k}=P^{\dag}_{\Psi_{k}}HP_{\Psi_{k}}$ with expectation value
$\varepsilon_{k}=\langle\Psi_{k},H_{k}\Psi_{k}\rangle$. This
observable corresponds to the concept of ``the energy of the
particle in the state $\Psi_{k}$''. This association of
observables with states does not violate the indistinguishability
of particles and the $N$ particle observable should be built as
the addition of all permutations of the tensor products as was
done for two particles in Eq.\ref{diffobserv}. Now the
generalization of Eq.\ref{id.dif} to $N$ particles tells us that
the correct treatment of $N$ identical particles in $N$ FAPP
differentiating states is equivalent to the treatment of $N$
different particles, each particle in one different state and
with its own values for its observables. But this is just the
classical system of $N$ particles.

There is however one case where the classical limit is not
defined (even at high temperature) and we must keep the correct
quantum mechanical treatment. Assume that we want to \emph{count}
the number of states that can be assumed by $N$ identical
particles compatible with some value of the total energy. This is
not an observable as before that could be associated to each
differentiating state. This number of states is not an
expectation value like the ones involved in Eq.\ref{id.dif} and
it would be therefore wrong to replace it by the number of states
obtained in the different particles case. The $N$ identical
particles are in one state among a large number of possibilities
and if we are interested in this number (and there are very good
reasons to be interested in it: the calculation of the entropy),
we have no way to relate it with some different particle
observable. We just must count the states for identical particles
taking care not to count twice the states that differ only by the
permutation of particles. The correct counting of states must
have the $N!$ introduced by Gibbs in order to obtain the correct
value for the entropy, without really knowing that its origin was
the quantum mechanical treatment of identical particles.

\section{EXAMPLES}
In order to consolidate the concepts presented in this work it is
convenient to apply them to some simple systems where, in some
cases, identical particles can be treated as different and in
some other cases not.

Consider the system of two hydrogen atoms located in space at
some distance $D$. When the distance is large compared with the
extension of the atoms, we can use the location in physical space
as FAPP differentiating states and we have the choice of treating
the system as composed by two different atoms (differentiated by
the location) or, on the contrary, to thereat it as two identical
atoms, as quantum mechanics dictates in rigour. When the distance
$D$ becomes smaller and smaller we reach the point where we no
longer have the choice and we must treat it with the correct
quantum mechanic recipe for identical atoms as is done when we
deal with the $H_{2}$ molecule. This treatment of the $H_{2}$
molecule can be found in many textbooks and will not be repeated
here but instead we can sketch how to deal with the two hydrogen
atom system when we are allowed to treat it as different atoms.
The one atom Hilbert space will involve two factors: one
describing the location of the atom in physical space, typically
the space of square integrable functions $\mathcal{L}_{2}$, and
the other factor involves the Hilbert space spanned by the energy
eigenstates of an electron in the Coulomb potential of the atom
that we denote by $\mathcal{H}_{E}$. A typical one atom state is
then an element of
$\mathcal{H}=\mathcal{L}_{2}\otimes\mathcal{H}_{E}$ like
$\Psi=\psi(x)\otimes\varphi_{n}$ (or $\Psi=(\psi,\varphi_{n})$ in
a simplified notation) where $\psi(x)$ describes the position of
the atom in physical space and $\varphi_{n}$ describes the
internal state of the atom, for instance, an energy eigenstate.
Similarly, another one atom state could be
$\Phi=\phi(x)\otimes\varphi_{m}=(\phi,\varphi_{m})$. Now, the two
atom state, considered as identical bosons, is described by a
state like $\frac{1}{\sqrt{2}}(\Psi\otimes\Phi+\Phi\otimes\Psi)$.
However if the states $\psi(x)$ and $\phi(x)$ are differentiating
(or FAPP differentiating) we may make the simplifying assumption
that the atoms are different and the state is $\Psi\otimes\Phi$.
In this case the differentiating criteria is ``the atom that is
located around the maximum of $|\psi(x)|^{2}$ for one of them and
around the maximum of $|\phi(x)|^{2}$ for the other'' (exactly
the same situation when I say ``the $Si$ atom in a chip of my
computer'' as different from the $Si$ atom in his computer). It
is important to emphasize however that the simplifying assumption
is not allowed when we want to know the number of states
consistent with some value of the energy. Perhaps nobody would be
interested in this number for our system of two atoms but for a
large number of atoms this number is needed in order to calculate
the entropy of the system. If we neglect the kinetic energy
associated with the movement of the atoms, the energy of the
system is determined by the two indices $(n,m)$ characterizing
the internal states of the atoms. Now, if the atoms \emph{were
distinguishable}, then we would have \emph{four states}
compatible with an energy given by $(n,m)$ (assuming for
simplicity that $n\neq m$). They are
$(\psi,\varphi_{n})\otimes(\phi,\varphi_{m})$,
$(\psi,\varphi_{m})\otimes(\phi,\varphi_{n})$,
$(\phi,\varphi_{n})\otimes(\psi,\varphi_{m})$ and
$(\phi,\varphi_{m})\otimes(\psi,\varphi_{n})$. However, the atoms
\emph{are identical} and some of theses states can not be
considered as different. Therefore we have only \emph{two states}
compatible with a given energy. They are
$\frac{1}{\sqrt{2}}[(\psi,\varphi_{n})\otimes(\phi,\varphi_{m})+
 (\phi,\varphi_{m})\otimes(\psi,\varphi_{n})]$ and
$\frac{1}{\sqrt{2}}[(\phi,\varphi_{n})\otimes(\psi,\varphi_{m})+
 (\psi,\varphi_{m})\otimes(\phi,\varphi_{n})]$.
This factor of two between the identical particle case and the
different particle case becomes the factor of $N!$ ``discovered''
by Gibbs, necessary for the entropy to be an extensive quantity.
Identical particles quantum mechanics gives the correct entropy
and the different particle approximation fails.

A system similar to the two hydrogen atoms is to consider two
identical particles placed un a double square well potential or
in two boxes. The different particle approximation when the two
square wells are widely separated is an interesting problem left
as an exercise. Notice that in this case, the energy eigenstates
are differentiating states (they ar orthogonal) but are not
localized in one or the other well. Adding and subtracting these
energy eigenstates we can build FAPP differentiating states
corresponding to placing a particle in one well or in the other.
\section{CONCLUSIONS}
In principle, the treatment of any particle by quantum mechanics
requires the symmetrization or anti-symmetrization of its state
with all other identical particles of its sort. Furthermore,
every particle or atom that we consider is just one representant
of myriads of other identical systems in the universe and
apparently a holistic treatment is necessary. However, we have
seen in this work that we may use the properties of some states
in order to separate out the system of interest from the rest and
consider it as a different particle or atom. We may therefore
think about ``this electron or this $Si$ atom right here'' as an
individual system differentiated from the rest by some property,
for instance, its localization within my computer. In this way,
the models that we build, for instance ``one single isolated
noninteracting hydrogen atom'', can be consider to be a faithful
representation of physical reality. In many cases, the treatment
of identical particles as different particles is FAPP justified
and may be more intuitive. However special care has to be taken
when we count the number of states associated with some value of
the energy in order to determine the entropy of the system
because the number of states in the different particle
approximation must be modified in order to get the correct
result.
\\ \\

One of us, H.O.M, acknowledges the ANPCyT (Argentina) for the
grant PICT 2004 Nr. 17-20075.

\end{document}